\documentclass[aps,twocolumn,showpacs,eqsecnum]{revtex4}
\usepackage{epsfig}
\usepackage{amsbsy}
\usepackage{amsmath}
\usepackage{amsfonts}

\begin{document}
\addtolength{\abovecaptionskip}{-2mm}
\addtolength{\belowcaptionskip}{-4mm}
\addtolength{\intextsep}{-6mm}
\date{\today}
\title{Phase measurements at the theoretical limit}
\author{D.\ W.\ Berry$^{1}$ and H.\ M.\ Wiseman$^{1,2}$}
\affiliation{$^{1}$Department of Physics, The University of Queensland,
St.\ Lucia 4072, Australia \\
$^{2}$School of Science, Griffith University, Nathan, Brisbane, Queensland 4111,
Australia}

\begin{abstract}
It is well known that the result of any phase measurement on an optical mode
made using linear optics has an introduced uncertainty in addition to the
intrinsic quantum phase uncertainty of the state of the mode. The best
previously published technique [H.\ M.\ Wiseman and R.\ B.\ Killip,
Phys.\ Rev.\ A
{\bf 57}, 2169 (1998)] is an adaptive technique that introduces a phase variance
that scales as $\bar{n}^{-1.5}$, where $\bar{n}$ is the mean photon number of
the state. This is far above the minimum intrinsic quantum phase variance of the
state, which scales as $\bar{n}^{-2}$. It has been shown that a lower limit to
the phase variance that is introduced scales as $\ln(\bar{n})/\bar{n}^2$. Here
we introduce an adaptive technique that attains this theoretical lower limit.
\end{abstract}
\maketitle

\newcommand{\nn}{\nonumber}
\newcommand{\nl}[1]{\nn \\ && {#1}\,}
\newcommand{\erf}[1]{Eq.\ (\ref{#1})}
\newcommand{\ip}[1]{\langle{#1}\rangle}
\newcommand{\bra}[1]{\langle{#1}|}
\newcommand{\ket}[1]{|{#1}\rangle}
\newcommand{\braket}[2]{\langle{#1}|{#2}\rangle}

\section{Introduction}
The phase of an electromagnetic field cannot be measured directly using
linear optics and photodetectors.
Rather than measuring phase directly, phase
measurement schemes rely on measuring quadratures of the field and inferring
the phase from these measurements. In a typical experimental
implementation, the
mode to be measured is passed through a 50/50 beam splitter, in order
to combine it with a much stronger local oscillator field. The difference
photocurrent from the two output ports of the beam splitter yields a
measurement of a particular quadrature.

The standard technique for measuring a completely unknown phase
is heterodyne detection, where all quadratures are
sampled with equal probability. This is achieved by using a local oscillator
field with a frequency slightly different from the signal's frequency, so its
phase changes linearly with respect to the phase of the signal. More accurate
phase measurements can be made using the homodyne technique, where the local
oscillator phase is $\Phi=\varphi + \pi/2$, with $\varphi$ the phase of the
signal. The problem with this is that it requires initial knowledge of the
phase of the signal, and so is not a phase measurement in the strict sense.

To maintain the unbiased nature of heterodyne phase measurements but obtain the
increased sensitivity of homodyne measurements, an adaptive dyne technique can
be used \cite{Wis95c,semiclass,fullquan,BerWisZha99}. Here ``dyne'' detection is
used to mean photodetection using a strong local oscillator at a beam splitter.
The idea behind adaptive phase measurement schemes is to use the information
gained so far during the measurement to estimate the system phase $\varphi$.
This is then used to adjust the local oscillator phase $\Phi$ to approximate a
homodyne measurement as above.

\begin{figure}[b]
\includegraphics[width=0.32\textwidth]{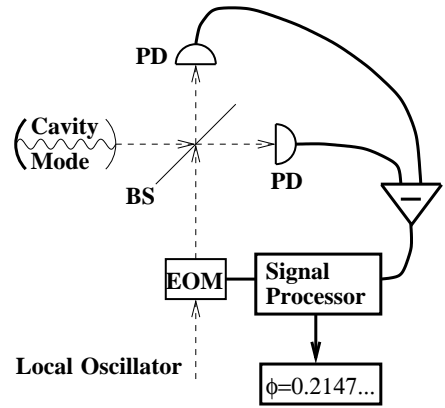}
\caption{Diagram of the apparatus for making an adaptive phase measurement.
The signal from the cavity is combined with the local oscillator field at a
50/50 beam splitter (BS) and the outputs are detected by photodetectors (PD).
The signals from these photodetectors are subtracted, and the difference signal
is processed by the digital signal processor, which determines a phase estimate
and adjusts the electro-optic phase modulator (EOM) accordingly.}
\label{apparatus}
\end{figure}

The apparatus for performing these measurements is shown schematically
in Fig.\ \ref{apparatus}. The signal and a local oscillator with
amplitude $\beta$ are combined at the beam splitter and the outputs are
measured with photodetectors. The outputs from the photodetectors,
$\delta N_+$ and $\delta N_-$, are subtracted and then fed into a digital signal
processor that uses these measurements to estimate the phase of the system,
and adjusts the phase of the local oscillator via an electro-optic phase
modulator. The signal is shown here as from a cavity with a half-silvered
mirror, as this is what is considered in the theory in Sec.\ \ref{method}.

The signal of interest is the difference between the photocurrents at the
two ports. We therefore define the signal as
\begin{equation}
\label{origdef}
I(t)\delta t=\frac{\delta N_+ -\delta N_-}{\beta e^{-t/2}}.
\end{equation}
Here we have used units of time such that the decay constant of the cavity
is unity. We divide by a factor of $e^{-t/2}$ because this is the square root of
the mode function for the signal. We can take account of signals with more
general mode functions $u(t)$ in a similar way \cite{semiclass}.

When we take the limit of very large local oscillator amplitude
and small time intervals $\delta t$, we find that
\begin{equation}
I(v)dv=2\text{Re}(\alpha^{S}_v e^{-i\Phi(v)})dv+dW(v),
\end{equation}
where $v$ is time scaled to the unit interval and $\alpha^{S}_v$ is the
scaled mean amplitude of the {\em system} (for which the $S$ superscript
stands). The systematic variation in the coherent amplitude with time due
to the mode shape is scaled out. This scaling is explained in more detail
in Sec.\ \ref{method}. The final term $dW(v)$ is an infinitesimal Wiener
increment such that $\langle{dW(v)^2}\rangle=dv$ \cite{text}.

It can be shown
\cite{wise96,fullquan} that just two complex numbers are necessary to
encapsulate all of the relevant information in the photocurrent record up to a
given time. These are
\begin{align}
A_v &= \int_0^v I(u)e^{i\Phi(u)}du ,\\
B_v &= -\int_0^v e^{2i\Phi(u)}du.
\end{align}
For convenience, we often replace $B_{v}$ by a third complex number defined
in terms of $A_{v}$ and $B_{v}$,
\begin{equation}
\label{defC}
C_v = A_v v+B_v A_v^*.
\end{equation}
Generally the best estimate of the
phase at time $v$ is $\arg(C_v)$ \cite{semiclass}.
The subscripts are omitted for the final values ($v=1$).

In adaptive measurement schemes the phase of the local oscillator is
generally taken
to be
\begin{equation}
\Phi(v) = \hat\varphi(v) + \frac \pi 2,
\end{equation}
where $\hat\varphi(v)$ is the estimated phase of the system at time
$v$ using the measurement results $A_{v}$ and $C_{v}$. There are a
number of possible phase estimates, giving different adaptive schemes.
For the mark I scheme \cite{semiclass,fullquan}, both
the running phase estimate $\hat\varphi(v)$
and the final phase estimate are taken to be $\arg(A_{v})$. This
is better than heterodyne measurements
only if the field is very weak \cite{Wis95c,fullquan}.
 For the mark II adaptive phase
measurements \cite{semiclass,fullquan} the best phase estimate
$\arg(C)$ is used at the end of the measurement, but for the
intermediate phase estimate $\arg(A_v)$ is used. This is better than
heterodyne measurements for all field strengths. If $\arg(C_{v})$ is
generally the best phase estimate, it is apparent from \erf{defC} that
$\arg(A_{v})$ will only be the best phase estimate if $B_v$ is negligible
(as it
is in the case of heterodyne measurements). For adaptive phase measurements
$B_{v}$ does not vanish and $\arg(A_{v})$ is generally a much worse phase
estimate than $\arg(C_v)$.

This raises the question of why this relatively
poor intermediate phase estimate is used.
There are two main reasons for this:
(i) it is possible to obtain direct analytic results for this case, whereas
using
a better intermediate phase estimate requires numerical evaluation;
(ii) the apparatus required to implement this method is much simpler than that
required for a better intermediate phase estimate.

Even with the relatively poor intermediate phase estimate,
 the mark II adaptive scheme introduces a phase variance of just
$\frac 18 (\bar n^{S})^{-1.5}$, a good improvement over the heterodyne
result of
$\frac 14 (\bar n^{S})^{-1}$. Here $\bar n^{S}$ is the mean photon number of
the field being measured, and the actual measured phase variance is
the introduced phase variance plus the intrinsic phase variance. The
intrinsic phase variance for a state of mean photon number $\bar n^{S}$
can be as small as of order $(\bar n^{S})^{-2}$ \cite{SumPeg90,BerWisZha99}.
This is far smaller than the introduced phase variance, so the latter
is what limits the accuracy of phase measurements.
Although the mark II results are far superior to the standard result
of heterodyne detection, it is still possible to improve on the mark II result,
and it is shown in Ref.\ \cite{fullquan} that a theoretical lower limit to
the phase variance that is introduced by an arbitrary phase measurement
scheme (based on linear optics and photodetection)
is $\frac 14 \ln (\bar n^{S})\times (\bar n^{S})^{-2}$.

In improving on the mark II result, the obvious thing to do is to use a better
intermediate phase estimate. It turns out that using the best phase estimate
$\arg(C_{v})$ actually gives a worse result than the mark II case, for
reasons that we will explain later. The phase estimates that
we consider in this paper are therefore intermediate between $\arg(A_v)$ and
the best phase estimate:
\begin{equation} \label{ifa}
\hat\varphi(v)=\arg(C_v^{1-\epsilon(v)} A_v^{\epsilon(v)}).
\end{equation}
It is possible to obtain a marked improvement over the mark II case by using
constant values of $\epsilon$. We show in Sec.\ \ref{results} that a scaling
of roughly $(\bar n^{S})^{-1.68}$ is possible. One drawback is that the
value of $\epsilon$ required depends on the photon number.

We can obtain an even better result if we allow $\epsilon$ to have a variation
in time, and we show in Sec.\ \ref{results} that we can obtain phase
estimates very close to the theoretical limit if we use
\begin{equation}
\epsilon(v) = \frac {v^2-|B_v|^2}{C_v} \sqrt{\frac v {1-v}}.
\end{equation}
This expression does not explicitly depend on the photon number. This method
works best if the phase estimates are updated in discrete time steps, and
the magnitude of the steps depends weakly on the photon number. A more serious
problem with this method is that it tends to produce values of $|B|$ that are
too close to 1. This means that final  phase estimates with an error close to
$\pi$ occur sufficiently frequently to make a significant contribution to the
phase uncertainty. We will show how this problem can be corrected.

The paper is structured as follows. In Sec.\ \ref{theory} we rederive the
ultimate theoretical limit to phase measurements of Ref.\ \cite{fullquan}. This
is necessary to understand how the improved feedback algorithm of \erf{ifa} can
approach the theoretical limit, as explained in Sec.\ \ref{impfeed}. In
Sec.\ \ref{method} we derive the results necessary for a numerical simulation of
this algorithm, and in Sec.\ \ref{results} present the results of those
simulations. The problem of infrequent results with large errors is identified
in Sec.\ \ref{evaluation} and a solution proposed and evaluated in
Sec.\ \ref{improved}. We conclude with a summary and discussion in
Sec.\ \ref{conclude}.

\section{The theoretical limit}
\label{theory}
In order to understand how to attain the theoretical limit, we must first
understand the reason for the theoretical limit. It can be shown
\cite{wise96} that the probability of obtaining the results $A$, $B$ from
an arbitrary (adaptive or nonadaptive) measurement is
\begin{equation}
P(A,B)d^{2}A\,d^{2}B = {\rm Tr}[\rho G(A,B)]d^{2}A\,d^{2}B,
\end{equation}
where $\rho$ is the state of the mode being measured. Here $G(A,B)$ is the POM
(probability operator measure) for the measurement, and is given by
\begin{equation}
G(A,B)=Q(A,B)\ket{\tilde\psi(A,B)}\bra{\tilde\psi(A,B)},
\end{equation}
where $Q(A,B)$ is what the probability distribution $P(A,B)$ would be if
$\rho$ were the vacuum state $\ket{0}\bra{0}$, and
$\ket{\tilde\psi(A,B)}$ is an unnormalized ket defined by
\begin{equation}
\ket{\tilde\psi(A,B)}=\exp\left[\frac{1}{2} B (a^\dagger)^2-A a^\dagger\right]
\ket{0}.
\end{equation}
This is proportional to a squeezed state \cite{WalMil94}:
\begin{equation}
\exp\!\!\left[\frac{1}{2} B (a^\dagger)^2\!-\!A a^\dagger\right]\!\!\ket{0}
\!=\!\left( 1\!-\!|B|^2
\right)^{-1/4}\!\exp(A{{\alpha}}^*\!/2)\ket{{{\alpha}},\xi},
\end{equation}
where
\begin{equation}
\ket{{{\alpha}},\xi}=\exp({{\alpha}} a^\dagger-{{\alpha}}^*a)
\exp\left[\frac{1}{2} \xi^* a^2-\frac{1}{2} \xi (a^\dagger)^2\right]\ket{0},
\end{equation}
and the squeezing parameters are
\begin{align}
\label{defineAB1}
\alpha &= \frac{A+BA^*}{1-|B|^2},\\
\label{defineAB2}
\xi &= -\frac{B \text{atanh}|B|}{|B|}.
\end{align}
where atanh is the inverse hyperbolic tan function.
In terms of these the POM is given by
\begin{equation}
G(A,B)=Q'(A,B)\ket{{{\alpha}},\xi}\bra{{{\alpha}},\xi},
\end{equation}
where
\begin{equation}
Q'(A,B)=Q(A,B)\left( 1-|B|^2 \right)^{-1/2} \exp\left[ \text{Re}\left(A
{{\alpha}}^*\right)\right].
\end{equation}
If the system state is pure, $\rho = \ket{\psi}\bra{\psi}$ and
the probability distribution is given by
\begin{equation}
P(A,B)=Q'(A,B)|\braket{\psi}{{{\alpha}},\xi}|^2.
\end{equation}

For an unbiased measurement scheme the probability distribution for the phase
resulting from this equation depends entirely on the inner product between the
two states, and not on $Q'(A,B)$. To see this, note first that if the
measurement is unbiased the vacuum probability distribution $Q(A,B)$ will be
independent of the phase. Second, for the squeezed state $\ket{\alpha,\xi}$,
$\xi \alpha^* /\alpha$ is independent of the phase $\arg(\alpha)$. This in
turn means that $BA^*/A$ is independent of the phase. Since
\begin{equation}
A\alpha^* = (1+BA^*/A)^* \frac{|A|^2}{1-|B|^2},
\end{equation}
$A\alpha^*$ and therefore $Q'(A,B)$ are independent of the phase.

Since the probability distribution for the phase depends on the inner product
between the two states, the variance in the measured phase will approximately
be the sum of the intrinsic phase variance and the phase variance of the
squeezed state $\ket{{{\alpha}},\xi}$. The maximum overlap between the states
will be when the squeezed state has about the same photon number as the input
state. This means that the theoretical limit to the phase variance that is
introduced by the measurement is the phase variance of the squeezed state that
has the same photon number as the input state and has been optimized for
minimum intrinsic phase variance. Since the phase variance of a squeezed state
optimized for minimum intrinsic phase variance is $\ln\bar n/(4\bar n^2)$ in the
limit of large $\bar n$ \cite{collett}, this is also the limit to the introduced
phase variance.

The photon number of the squeezed state at maximum overlap will be mainly
determined by the photon number of the input, but the degree and
direction of squeezing (parametrized by $\xi$) will be
determined by the multiplying factor $Q'(A,B)$. The multiplying factor can be
expressed as a function of $\bar n$ and $\zeta$, for which we will
use the same symbol $Q'$, even though it is a new function
$Q'(\bar{n},\zeta)$. Here  $\bar n$ is
the mean photon number for the state $\ket{{{\alpha}},\xi}$ (and will be close
to the photon number $\bar{n}^{S}$ of the input state), and $\zeta = \xi
\alpha^* /\alpha$ is $\xi$ with the phase of $\alpha$ scaled out. The
multiplying factor will tend to be concentrated along a particular line,
effectively giving $\zeta$ as a function of $\bar n$.
 In order to obtain the theoretical limit, the measurement
scheme must give a multiplying factor $Q'(\bar n,\zeta)$ that tends to give
values of $\zeta$ for each $\bar n$ that are the same as for optimized squeezed
states.

We can determine the approximate variation of $\zeta$ with $\bar n$ in the
multiplying factor if we can estimate how it varies for measurements on a
coherent state. If we consider measurements on a coherent state with real
amplitude $\alpha^{S}$, then the maximum overlap with the state
$\ket{\alpha,\xi}$ will be for $\alpha^{S} \approx {{\alpha}}$. We use
$\alpha^S$ without a subscript to indicate the initial coherent amplitude
before the measurement.

If we are using an adaptive scheme with intermediate phase estimates that are
unbiased, it is easy to see that the maximum probability will be for $B$ real
and therefore also $A$ real. These results imply that
\begin{equation}
{{\alpha}} \approx \frac{A(1+B)}{1-B^2} = \frac{A}{1-B}.
\end{equation}
In turn this gives $\zeta$ as
\begin{align}
\zeta &\approx -\text{atanh}(1- A/{{\alpha}}) \\
&\approx \frac 1 2 \ln \frac A{2{{\alpha}}} \approx \frac 1 2 \ln \frac
{A}{2\sqrt{{{\bar n}}}}.
\label{defzetaAn}
\end{align}
Since the value of $\zeta$ is governed by the multiplying factor
$Q'(\bar n,\zeta)$, this result for $\zeta$ should hold for more general input
states.

From Ref.\ \cite{collett} the phase variance of a squeezed state is
\begin{equation} \label{pvss}
\langle\Delta \phi^2\rangle\approx\frac{n_0+1}{4\bar n^2}+2\text{erfc}
(\sqrt{2n_0}).
\end{equation}
where $n_0=\bar n e^{2\zeta}$ for real $\zeta$. This is minimized
asymptotically as
\begin{equation}
\frac{\ln\bar{n} + \Delta}{4\bar n^2},
\end{equation}
where $\Delta \approx 2.43$, for
\begin{equation}
n_0 \approx \ln(4{{\bar n}})-\frac 14 \ln(2\pi).
\end{equation}
If we use the result obtained for $\zeta$ in Eq.\ (\ref{defzetaAn}) we find that
\begin{equation}
n_0 \approx \frac 1 2 |A| \sqrt{{{\bar n}}}.
\label{n0result}
\end{equation}
This result means that in order for the measurement to be optimal, $|A|$ should
scale with ${{\bar n}}$ as
\begin{equation}
|A| \propto \frac{\ln {{\bar n}}}{\sqrt{{{\bar n}}}}.
\end{equation}
For the case of mark II measurements we have the result that $|A|=1$
\cite{semiclass}, which is why these measurements are not optimal. Note that if
we substitute $|A|=1$ into the expression (\ref{n0result}) to find
$n_{0}$, and substitute that into \erf{pvss}, we obtain the correct result for
the mark II introduced phase variance,
\begin{equation}
\langle{\Delta \phi^2}\rangle\approx \frac 18 {{\bar n}}^{-1.5}.
\end{equation}

\section{Improved feedback}
\label{impfeed}
Now we have the result that for optimal feedback $|A|$ should decrease with
photon number. Therefore in order to improve the phase measurement scheme we
want one that gives $|A|<1$. To see in general how this can be achieved,
consider a coherent state with amplitude $\alpha^{S}$ and determine the Ito
SDE (stochastic differential equation) for $|A|^2$:
\begin{align}
d|A_v|^2&=A_v^*(dA_v)+(dA_v^*)A_v+(dA_v^*)(dA_v) \\
&=A_v^* e^{i\Phi(v)}I(v)dv + e^{-i\Phi(v)}I(v)dv A_v +dv \\
&=[ |A_v| I(v) 2{\rm Re}( e^{i\Phi(v)}e^{-i\varphi_v^A})
+1] dv,
\end{align}
where $\varphi_v^A=\arg A_v$. In terms of the phase estimate
$\hat \varphi_v = \Phi(v)-\pi/2$ this becomes
\begin{equation}
d|A_v|^2=[1+ 2|A_v| I(v) \sin(\varphi_v^A-\hat\varphi_v)] dv.
\end{equation}
If we take the expectation value of $I(v)$ and simplify we get
\begin{equation}
\langle{I(v)}\rangle=-2|\alpha^{S}|\sin(\hat \varphi_v-\varphi),
\end{equation}
where $\varphi=\arg \alpha^{S}$.
If we use this result the expectation value for the increment in $|A_v|^2$ is
\begin{equation}
\label{sines}
\ip{d|A_v|^2}=\left[1- 4|A_v||\alpha^{S}|\sin(\hat \varphi_v-\varphi)
\sin(\varphi_v^A-\hat\varphi_v) \right] dv.
\end{equation}
The first term on its own will give $|A|=1$, and in order to get $|A|<1$
the two
sines must have the same sign. This will be the case if the phase estimate is
between the actual phase and the phase of $A_v$. It is for this reason that we
consider phase estimates that are intermediate between the best phase estimate
and the phase of $A_v$, i.e., of the form
\begin{equation}
\hat\varphi(v)=\arg[(A_v v+B_v A_v^*)^{1-\epsilon(v)}A_v^{\epsilon(v)}].
\end{equation}
In general, smaller values of $|A|$ can be obtained by using smaller values of
$\epsilon$. This is because $\varphi_v^A$ tends to be a worse phase estimate,
thus making it possible for the sines in \erf{sines} to be larger.
Note that it is far too simplistic to use the best phase estimate (i.e., with
$\epsilon = 0$), as we need to adjust $\epsilon$ in order to make $n_0$ closer
to optimal.

\section{Simulation method}
\label{method}
The easiest input states to use for numerical simulations are
coherent states, as they remain coherent with a deterministically
decaying amplitude. However,
in order to estimate the phase variance that is introduced by the measurement
this would be very inefficient, as the phase
variance would be dominated by the intrinsic phase variance. It is
almost as easy (and much more efficient) to
perform calculations on squeezed states, as squeezed states remain
squeezed states under the stochastic evolution, and only the two squeezing
parameters need be kept track of. The best squeezed states to use are those
optimized for minimum intrinsic phase variance. For these states the total
phase variance will be approximately twice the intrinsic phase variance
when the measurements are close to optimal.

To determine the SDE's for the squeezing parameters, we must first consider the
SDE for the state. For dyne detection the stochastic evolution of the
conditioned state vector is \cite{wise96}
\begin{align}
d\ket{\psi (t)} &=\left[ dt \left( \frac{\ip{a^\dagger a}}2
-\frac{a^\dagger a}2 +
\frac{\ip{a^\dagger\gamma+\gamma^*a}}2 -\gamma^*a\right)\right. \nn \\
&~~~\left. + dN(t)\left( \frac{a e^{-i\Phi}+|\gamma|}
{\sqrt{\ip{(a^\dagger+\gamma^*)(a+\gamma)}}} -1\right)
 \right] \ket{\psi(t)}, \nn \\ \label{thiseq}
\end{align}
where $a$ is the annihilation operator for the mode, $|\gamma|\gg 1$ is the
amplitude of the local oscillator, and $\Phi=\arg{\gamma}$ is its phase.
Here the mode being measured is assumed to come from a cavity with an
intensity decay rate equal to unity. The
point process $dN(t)$ has a mean $\kappa dt$, where
\begin{equation}
\kappa=\langle{(a^\dagger+\gamma^*)(a+\gamma)}\rangle.
\end{equation}
The equation given in
\cite{wise96} differs from \erf{thiseq} by a trivial phase factor. The form
above is given because it is not possible to directly take the limit of large
local oscillator amplitude using the form given in \cite{wise96}. To take the
limit of large local oscillator amplitude we approximate the Poisson process
$\delta N(t)$ by a Gaussian process
\begin{equation}
\delta N(t) \approx \kappa \delta t+\sqrt{\kappa}\delta W(t),
\end{equation}
where $\delta W(t)$ is a Gaussian random variable of zero mean and variance
$\delta t$. Then we find that in the limit of large $|\gamma|$ we have
\begin{align}
d\ket{\psi(t)} &=[(-{a^\dagger a}/2+a\chi e^{-i\Phi}-{\chi^2}/2)dt \nn \\ & ~~~+
(a e^{-i\Phi}-\chi)dW] \ket{\psi(t)},
\end{align}
where
\begin{equation}
\chi = \frac 12 (\langle{a}\rangle e^{-i\Phi}+
\langle{a^\dagger}\rangle e^{i\Phi}).
\end{equation}

In order to determine the SDE's for the squeezing parameters, we use the
method of Rigo {\it et al.}\ \cite{rigo}. Squeezed states obey the relation
\begin{equation}
( a-B_t^S a^\dagger -A_t^S ) \ket{A_t^S,B_t^S}.
\end{equation}
The squeezing parameters $A_t^S$ and $B_t^S$ are related to the usual squeezing
parameters in the same way as $A$ and $B$ are in \erf{defineAB1} and
\erf{defineAB2}. In the Stratonovich formalism
\begin{equation} \label{LHSRHS}
(a-B_t^S a^\dagger-A_t^S)d\ket{\psi(t)}=(dB_t^Sa^\dagger+dA_t^S)\ket{\psi(t)}.
\end{equation}
Converting the SDE for the state to the Stratonovich form in the usual way
\cite{text}, we find
\begin{align}
d\ket{\psi(t)} &= \left[ \left( -\frac{a^\dagger a}2-\frac{a^2
e^{-2i\Phi}}2 + 2a
\chi e^{-i\Phi} -\chi^2 \right) dt \right.
\nn \\ & ~~~+( ae^{-i\Phi} -\chi ) dW
\nn \\ & ~~~\left. -\frac 12 [ ad(e^{-i\Phi})-d\chi ] dW \right]\ket{\psi(t)}.
\end{align}
Here we have included the increments $d(e^{-i\Phi})$ and $d\chi$
because the phase of the
local oscillator can vary stochastically. Using this form of the
equation, the left hand side of \erf{LHSRHS} evaluates to
\begin{align}
&\left\{ \vphantom{\frac 12} \!
dt[-(a^\dagger B_t^S\!+\!A_t^S/2)\!-\!B_t^S(B_t^Sa^\dagger\!+\!A_t^S)
e^{-2i\Phi}\!\!+\!2B_t^S\chi e^{-i\Phi}]\right. \nn \\ & ~~~~\left. 
+dW\left[B_t^S e^{-i\Phi} - \frac 12 A_t^S d(e^{-i\Phi}) \right] \right\}
\ket{\psi(t)}. \nn
\end{align}
This gives us the SDE's for the squeezing parameters,
\begin{align}
dB_t^S &= -B_t^S(1+e^{-2i\Phi}B_t^S) dt, \\
dA_t^S &= -\frac 12 A_t^Sdt\!+\!B_t^S\langle{a^\dagger}\rangle (1\!+\!
e^{-2i\Phi}B_t^S)dt\!+\!B_t^S e^{-i\Phi}dW \nn \\
&~~~-\frac 12 B_t^S d(e^{-i\Phi}) dW.
\end{align}

From these we find that the Stratonovich SDE for the standard (nonscaled)
amplitude $\alpha^{S}_t$ is
\begin{align}
d\alpha^{S}_t &= -\frac 12 \alpha^{S}_t dt+\frac{B_t^S dW}{1-|B_t^S|^2}
[(B_t^S)^*e^{i\Phi}+e^{-i\Phi}] \nn\\ &~~~-\frac 12 \frac{B_t^S dW}{1-|B_t^S|^2}
[(B_t^S)^* d(e^{i\Phi})+d(e^{-i\Phi})]. \nn \\
\end{align}
Converting back to the Ito SDE, we get
\begin{align}
d\alpha^{S}_t  = -\frac 12 \alpha^{S}_t  dt +\frac{B_t^S
dW}{1-|B_t^S|^2}[(B_t^S)^* e^{i\Phi}+e^{-i\Phi}]. \nn \\
\end{align}
The SDE for $B_t^S$ is unchanged under the change to Ito form. If we take
the signal to be $I(t)\delta t=(\delta N_+ -\delta N_-)/\beta$ (for
consistency with Ref.\ \cite{wise96}), then take the limit of large oscillator
amplitude and small time intervals $\delta t$, we obtain
\begin{equation}
I(t)dt=2\text{Re}(\alpha^{S}_t  e^{-i\Phi(t)})dt+dW(t).
\end{equation}
The parameters $A_t$ and $B_t$ are then defined as in \cite{wise96} by
\begin{align}
A_t &= \int_0^t e^{i\Phi}e^{-s/2}I(s)ds, \\
B_t &= - \int_0^t e^{2i\Phi} e^{-s} ds.
\end{align}
In order to get rid of the exponential factors, we change the time variable to
\begin{equation}
v=1-e^{-t},
\end{equation}
and we redefine the amplitude to remove the systematic variation:
\begin{equation}
\label{scaled}
\alpha^{S}_v = \alpha^{S}_t e^{t/2}.
\end{equation}
Here we use the $v$ subscript to indicate the scaled amplitude, and the $t$
subscript to indicate the original, unscaled amplitude. Since these are equal
to each other at zero time, there is no ambiguity in the initial amplitude
$\alpha^S$. Reverting to our original definition of the signal
(\ref{origdef}), we find
\begin{equation}
I(v)dv = 2\text{Re}(\alpha^{S}_v e^{-i\Phi(v)})dv+dW(v).
\end{equation}
With these changes of variables, the definitions for $A_v$ and $B_v$ become
\begin{align}
A_v &= \int_0^v e^{i\Phi}I(u)du, \\
B_v &= - \int_0^v e^{2i\Phi} du.
\end{align}
The differential equations for the squeezing parameters become
\begin{align}
\label{squdeq}
dB_v^S &= -\frac{dv}{1-v}B_v^S (1+e^{-2i\Phi}B_v^S), \\
d\alpha^{S}_v &= \frac 1{1-v} \frac{B_v^S dW(v)}{1-|B_v^S|^2}[ (B_v^S)^*
e^{i\Phi} +e^{-i\Phi} ].
\end{align}
Initial calculations were performed using these equations, but there is a
further simplification that can be made. The solution for $B_v^S$ is
\begin{equation}
B_v^S = \frac {1-v}{(B_0^S)^{-1}-B_v^*}.
\end{equation}
For calculations with time-dependent $\epsilon$ this solution for $B_v^S$
was used rather than solving a separate differential equation for $B_v^S$.

\section{Results}
\label{results}
\begin{figure}[t]
\includegraphics[width=0.45\textwidth]{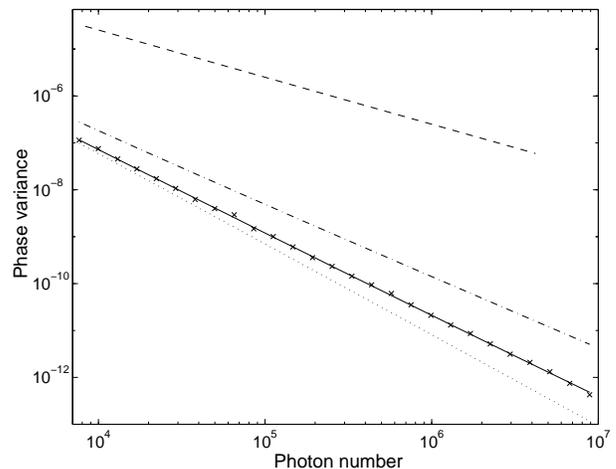}
\caption{Phase variance for phase measurements with a constant value of
$\epsilon$ plotted as a function of the photon number of the input state. The
crosses are the values obtained by stochastic integration and the continuous
line is the fitted line. For comparison we have also plotted, in order from top
to bottom, the variance for heterodyne measurements (dashed line), for mark II
measurements (dash-dotted line), and the theoretical limit (dotted line).}
\label{constant}
\end{figure}

First we will describe the results for constant $\epsilon$. For each mean
photon number, $\epsilon$ was varied to find the value that gave the minimum
phase variance. This method does not give results close to the theoretical
limit for photon numbers above about 5000, but the phase variances continue to
get smaller as compared to the phase variances for mark II measurements. This
indicates that the results are following a different scaling law, and fitting
techniques give the power for the introduced phase variance as $1.685\pm
0.007$. The data and the fitted line along with the heterodyne and mark II cases
and the theoretical limit are shown in Fig~\ref{constant}. These results are
a significant improvement over the mark II case, but are still significantly
above the theoretical limit.

In order to improve on this result we must vary $\epsilon$ during the
measurement. The value of $\epsilon$ that we found to give the best result was
\begin{equation}
\label{epval}
\epsilon(v) = \frac {v^2-|B_v|^2}{|C_v|} \sqrt{\frac v {1-v}}.
\end{equation}
The reason for the multiplying factor of $(v^2-|B_v|^2)/|C_v|$ is that it is an
estimator for $1/|\alpha^{S}|$. This means that the value of $\epsilon$
tends to be
smaller for larger photon numbers, resulting in smaller values of $|A|$. The
reason for the factor of $\sqrt{v/(1-v)}$ is that it makes the value of
$\epsilon$ close to zero initially, and very large near the end of the
measurement.

This second factor was found essentially by trial and error, and is thought
to be related to the fact that the phase of $\alpha^{S}_v$ varies stochastically
during the measurement. Recall that during the measurement we want the phase
estimate to be between the phase of $\alpha^{S}_v$ and the phase of $A_v$. We
only have an estimate of the phase of $\alpha^{S}$ (the initial phase), so if
we use a phase estimate that is too close to the actual phase when the phase
variance of $\alpha^{S}_v$ is large, the phase estimate is likely to be outside
the interval between the phase of $\alpha^{S}_v$ and the phase of $A_v$.
Since the phase variance of $\alpha^{S}_v$ increases with time, the value of
$\epsilon$ is increased as well, to prevent this happening.

\begin{figure}[b]
\includegraphics[width=0.45\textwidth]{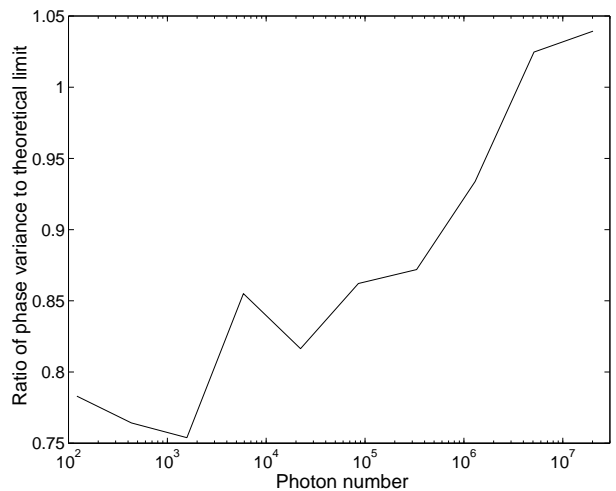}
\caption{Phase variance for phase measurements with a time-dependent $\epsilon$
plotted as a function of the photon number of the input state. The phase
variance is plotted as a ratio to the theoretical minimum phase variance (i.e.,
twice the intrinsic phase variance).}
\label{ratio}
\end{figure}

The results for this method are shown in Fig.\ \ref{ratio} as a ratio to the
theoretical limit. As this shows, the results are very close to the theoretical
limit, and even for the largest photon number for which calculations have been
performed the phase uncertainty is only about 4\% above the theoretical limit.
For these calculations the time steps used were approximately
\begin{equation}
\Delta v=\frac{\bar n^{S} \langle{\Delta \phi^2}\rangle_{\rm th}}{25},
\end{equation}
where $\langle{\Delta \phi^2}\rangle_{\rm th}$ is the theoretical limit to the
phase uncertainty. With these time steps the uncertainty due to the finite step
size is approximately 1\%.

If the integration time step is reduced, while keeping the time interval at
which the phase estimates are updated constant, the phase variance converges.
If, however, the phase estimates are updated at smaller and smaller time
intervals then the phase variance does not converge. For example, the phase
uncertainty for measurements on an optimized squeezed state with a photon
number of 1577 is $1.54\times 10^{-6}$ if we use the time steps given above.
If, however, we use time steps that are 100 times smaller, then the phase
variance is $1.93\times 10^{-6}$, and if the time steps are 1000 times
smaller the phase variance is $2.13\times 10^{-6}$. These results indicate that
the phase estimates must be incremented in finite time intervals for this
method to give good results, and the size of the time steps that should be used
depends on the photon number. The phase variance is not strongly dependent on
these time steps, however, and only an order of magnitude estimate of the
photon number is required.

\section{Evaluation of method}
\label{evaluation}
A problem with determing the phase variance by the method above is that, for
highly squeezed states (that are close to optimized for minimum phase
variance), a significant contribution to the phase variance is from low
probability results around $\pi$. In obtaining numerical results the actual
phase variance for the measurement will tend to be underestimated because the
results from around $\pi$ are obtained too rarely for good statistics. It
would require an extremely large number of samples to estimate this
contribution. However, we can estimate it nonstatistically as follows.

\begin{figure}
\includegraphics[width=0.45\textwidth]{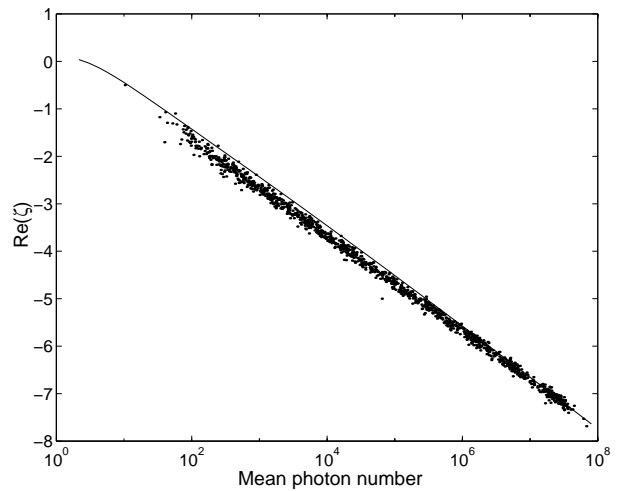}
\caption{Values of $\zeta_R$ and $\bar n$ (calculated from $A$ and $B$)
resulting from measurements on squeezed states of various mean photon numbers.
The variation of $\zeta$ with $\bar n$ for optimum squeezed states is also
plotted (continuous line).}
\label{zeta}
\end{figure}

Recall that in order to have a measurement that is close to optimum the
multiplying factor $Q'(\bar n,\zeta)$ should give values of $\zeta$ for each
$\bar n$ that are close to optimized for minimum phase uncertainty. To test this
for the phase measurement scheme described above, the $\bar n$ and $\zeta$ were
determined from the values of $A$ and $B$ from the samples. The resulting data
along with the line for optimized $\zeta$ are plotted in Fig.\ \ref{zeta}. The
imaginary part of $\zeta$ should be zero for optimum measurements, and is small
for these results. Therefore in Fig.\ \ref{zeta} we have plotted the real part
$\zeta_R$. As can be seen, the vast majority of the data points are below the
line, indicating greater squeezing than optimum. This means that if the low
probability results around $\pi$ are taken into account the phase variance for
these measurements will be above the theoretical limit.

First we consider the effect of variations in the modulus of $\zeta$, leaving
consideration of error in the phase till later. In order to estimate how far
above the theoretical limit the actual phase variance is, we make a quadratic
approximation to the expression for the phase variance. From \cite{collett} the
expression for the phase variance of a squeezed state is, for real $\zeta$,
\begin{equation}
\langle{\delta\phi^2}\rangle \approx \frac{e^{2\zeta}}{4\bar n}+\frac 1
{4\bar n^2}2 \text{erfc}(\sqrt{2\bar n}e^\zeta).
\end{equation}
Taking the derivative with respect to $\zeta$ gives
\begin{equation}
\frac d{d\zeta}\langle{\delta\phi^2}\rangle \approx \frac{e^{2\zeta}}{2\bar n}
-4e^\zeta\sqrt{\frac{2\bar n}{\pi}}e^{-2\bar n e^{2\zeta}}.
\end{equation}
Taking the second derivative and using the fact that the expression above is
zero for minimum phase variance gives
\begin{equation}
\frac {d^2}{d\zeta^2}\langle{\delta\phi^2}\rangle \approx \frac{n_0}
{2\bar n^2}(1+4n_0).
\end{equation}
This means that for values of $\zeta$ close to optimum the increase in the
phase variance over the optimum value is
\begin{equation}
\Delta \langle{\delta\phi^2}\rangle \approx (\Delta |\zeta|)^2 \frac{n_0}
{4\bar n^2}(1+4n_0).
\end{equation}
The main contribution to the phase uncertainty is $n_0/(4\bar n^2)$, so
the increase in the phase uncertainty as a ratio to the minimum phase
uncertainty is
\begin{equation}
\frac{\Delta \langle{\delta\phi^2}\rangle}{\langle{\delta\phi^2}
\rangle_{\rm min}} \approx (\Delta |\zeta|)^2 (1+4n_0).
\end{equation}
This estimate indicates that the actual phase variance for the measurement
scheme described above can be significantly larger than the intrinsic phase
variance. For example, for a mean photon number of about 332$\,$000 the rms

deviation of $|\zeta|$ from the optimum value is only about 0.16, but a
squeezed state with $|\zeta|$ differing this much from optimum will have a
phase variance more than twice the optimum value. This indicates that if the
low probability results around $\pi$ are taken into account the introduced
phase variance is actually more than twice the theoretical limit.

Next, we estimate the contribution from error in the phase (rather than the
modulus) of $\zeta$. For a squeezed state with real $\alpha$ the intrinsic
uncertainty in the zero quadrature is
\begin{equation}
\ip{X_0^2}=e^{-2|\zeta|}\cos^2 \frac{\mu}2+e^{2|\zeta|}\sin^2 \frac{\mu}2,
\end{equation}
where $\mu = \arg \zeta$. Since $X_0=2\alpha \sin(\phi)\approx 2
\alpha \phi$, the intrinsic uncertainty in the phase is
\begin{equation}
\ip{\delta\phi^2} \approx \frac{e^{-2|\zeta|}\cos^2(\mu/2)+e^{2|\zeta|}
\sin^2(\mu/2)}{4\bar n}.
\end{equation}
If the phase of $\zeta$ is small, we can make the approximation
\begin{equation}
\ip{\delta\phi^2} \approx \frac{e^{-2|\zeta|}+e^{2|\zeta|}\mu^2/4}{4
\bar n}.
\end{equation}
Clearly the first term in the numerator is just the original phase variance,
and the second term is the excess phase variance due to the error in the phase
of $\zeta$. Therefore the extra phase variance due to error in the phase of
$\zeta$ is given by
\begin{equation}
\Delta \ip{\delta\phi^2} \approx \frac{(\Delta \arg \zeta)^2}{16 n_0}.
\end{equation}

Using this estimate on the previous example it can be seen that this is not so
much of a problem, with the introduced phase uncertainty being increased by
less than 3\% by this factor.

\section{Improved method}
\label{improved}
The problem of the large contribution of the low probability results around
$\pi$ can be effectively eliminated in the following way. At each time step the
photon number is estimated from the values of $A_v$ and $B_v$, and the optimum
value of $\zeta$ is estimated using the asymptotic formula in \cite{collett}.
Then if $\zeta_R$ (the real part of $\zeta$) is too far below the optimum
value, rather than using the feedback phase above, we use
\begin{equation}
\label{altfeed}
\Phi(v) = \frac 12 \arg \left[ B_v-v\frac{C_v}{C_v^*} \tanh \left|
\zeta_{\rm opt}\right|\right].
\end{equation}
Using this feedback phase takes $B_v$ directly towards the optimum value. To see
this, note that the optimum value of $B_v$ is
\begin{equation}
B_v^{\rm opt} = v\frac{C_v}{C_v^*}\tanh \left|\zeta_{\rm opt}\right|.
\end{equation}
Taking the exponential of the feedback phase given by Eq.\ (\ref{altfeed}) gives
$e^{2i\Phi}\propto B_v-B_v^{\rm opt}$, so $dB_v\propto B_v^{\rm opt}-B_v$.

The details of exactly when $\zeta_R$ is considered too far below optimum can
be varied endlessly, but for the results that will be presented here we use
this alternate phase estimate after time $v=0.9$ and when
\begin{equation}
\label{toohi}
|\zeta|>|\zeta_{\rm opt}|e^{\lambda|\alpha_v|^2(1-v)},
\end{equation}
where $\zeta_{\rm opt}$ is the estimated optimum value of $\zeta$ and $\alpha_v$
is $C_v/(v^2-|B_v|^2)$. Using the exponential multiplying factor means that the
alternative feedback is used only toward the end of the measurement. Only
considering the alternative feedback in the last 10\% of the
measurement is necessary for the smaller photon numbers, where
Eq.\ (\ref{toohi}) is too weak a restriction.

Another variation from the previous scheme is that, for the larger photon
numbers, the values of $\epsilon$ given by the original expression were reduced.
The above correction corrects only for values of $\zeta_R$ that are below
optimum, and for the larger photon numbers many of the uncorrected values of
$\zeta_R$ tend to be above optimum (see Fig.\ \ref{zeta}). The corrections will
still work well, however, if we use a dividing factor to bring the uncorrected
values below the line. For the second largest photon number tested of around
$5\times 10^5$, the best results were obtained when the values of $\epsilon$ as
given by Eq.\ (\ref{epval}) were divided by 1.1. For the largest mean photon
number tested, $2\times 10^7$, the best results were obtained for a dividing
factor of 1.2. The value of $\lambda$ that gave the best results with these
dividing factors was $5\times 10^{-4}$. For all other mean photon numbers the
value of $\lambda$ used was $10^{-3}$.

\begin{figure}
\includegraphics[width=0.45\textwidth]{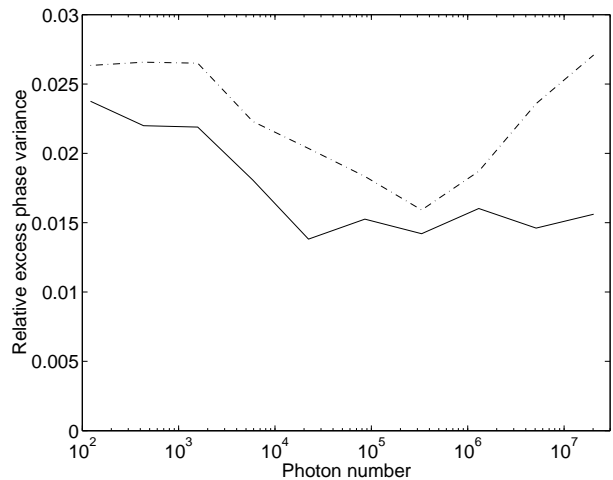}
\caption{Contributions to the phase uncertainty from error in the magnitude of
$\zeta$ (continuous line) and the phase of $\zeta$ (dash-dotted line). Dividing
factors of $1.1$ and $1.2$ are used for photon numbers of $5\times 10^6$ and
$2\times 10^7$, respectively. For these photon numbers $\lambda=5\times
10^{-4}$, otherwise $\lambda=10^{-3}$. The contributions are plotted as a ratio
to the theoretical minimum introduced phase uncertainty.}
\label{contrib}
\end{figure}

The estimated contributions to the phase variance due to error in the magnitude
and phase of $\zeta$ are plotted in Fig.\ \ref{contrib}. As can be seen, the
contribution due to error in the magnitude of $\zeta$ is very small, around
1.5\% for the larger photon numbers tested. The contribution due to the error in
the phase of $\zeta$ is a bit larger, but it still does not rise above 3\%.
Thus we can see that the introduced phase variance can be made very close
to the theoretical limit, within 5\% for the largest photon number tested.

With this modified technique the phase variance again does not converge as the
feedback phase is updated in smaller and smaller time intervals. The phase
variance is less dependent on the time step with this technique, however. For
example, for a mean photon number of 1577 the total phase variance for
measurements on an optimized squeezed state increases by only about 9\% as the
time steps are reduced by a factor of 1000. In contrast, the phase variance
increases by a factor of 38\% for the previous technique.

\section{Conclusions}
\label{conclude}

Any estimate of an initially unknown optical phase made using standard devices
(linear optical and opto-electronic devices, a local oscillator, and
photodetectors) must have an uncertainty above the intrinsic quantum uncertainty
in the phase of the input state. The minimum magnitude of the added phase
variance was determined in Ref.\ \cite{fullquan} to scale asymptotically as
\begin{equation}
\label{fund}
\frac{\ln\bar{n}^{S}}{4(\bar{n}^{S})^{2}},
\end{equation}
where $\bar{n}^{S}$ is the mean photon number of the input state. Previous
phase
measurement schemes do not approach this theoretical limit. In this paper we
have shown that an adaptive phase measurement scheme not previously considered
can attain this theoretical limit. In other words, we have determined what is
essentially the best possible phase measurement technique.

In practice, phase measurements are currently limited by detector
inefficiency. For detector efficiency $\eta$ the introduced phase variance
cannot be reduced below \cite{semiclass}
\begin{equation}
\label{effic}
\frac{1-\eta}{4\eta \bar{n}^{S}}.
\end{equation}
When the mark II phase variance is less than this there is not likely to be
any significant advantage to using a more advanced feedback scheme. For
the best photodetectors available today, with around 98\% efficiency
\cite{polzik}, the mark II phase variance falls below this limit for photon
numbers above 1000. Below this photon number the mark II phase variance is
never more than about 27\% above the limits determined using Eqs.\ (\ref{fund})
and (\ref{effic}), so only relatively small
improvements can be obtained by using a
more advanced feedback scheme.

Nevertheless, the technology is always improving, and there is no fundamental
reason why photodetectors cannot be built with efficiencies extremely close to 1
\cite{hidpc}. When very efficient photodetectors are developed, the feedback
techniques described here have the potential to give great improvements in the
accuracy of phase measurements for applications where there is a limitation on
the photon number that can be used. The other detrimental factors are relatively
minor, although the time delay in the feedback loop will become significant for
very short pulses.

The primary significance of the result obtained in this paper is theoretical,
however, as it represents the culmination of the search for the best optical
phase measurement schemes using standard devices. To do any better would require
using nonlinear optical devices. For example, it is conceivable that
down-converting some portion of the signal field, and then measuring the phase
of the down-converted light, could enable the above theoretical limit to be
surpassed. This is a question for future work.

\end{document}